\documentclass[prb,twocolumn,superscriptaddress,showpacs,floatfix]{revtex4}

\newcommand{\ket}[1]{\ensuremath{\left|{#1}\right\rangle}}

\usepackage[sort&compress]{natbib}
\usepackage{graphicx, times}
\usepackage{color}

\begin{document}
\title{Perfect Microwave Photodetection in Circuit QED}
\author{B. Peropadre}
\address{Instituto de F{\'\i}sica Fundamental, CSIC, Calle Serrano 113-bis, Madrid
E-28006, Spain}
\author{G. Romero}
\address{Departamento de Qu{\'\i}Â­mica F{\'\i}sica, Universidad del Pa{\'\i}s Vasco - Euskal Herriko Unibertsitatea, Apdo. 644, 48080 Bilbao, Spain}
\author{G. Johansson}
\address{Department of Microtechnology and Nanoscience, Chalmers University of Technology, G\"oteborg, Sweden}
\author{C. M. Wilson}
\address{Department of Microtechnology and Nanoscience, Chalmers University of Technology, G\"oteborg, Sweden}
\author{E. Solano}
\address{Departamento de Qu{\'\i}Â­mica F{\'\i}sica, Universidad del Pa{\'\i}s Vasco - Euskal Herriko Unibertsitatea, Apdo. 644, 48080 Bilbao, Spain}
\affiliation{IKERBASQUE, Basque Foundation for Science, Alameda Urquijo 36, 48011 Bilbao, Spain}
\author{J.~J. Garc{\'\i}a-Ripoll}
\address{Instituto de F{\'\i}sica Fundamental, CSIC, Calle Serrano 113-bis, Madrid
E-28006, Spain}
\date{\today}

\begin{abstract}
In this work we propose a microwave photon detector which successfully reaches 100\% efficiency with only one absorber. Our design consists of a metastable quantum circuit coupled to a semi-infinite transmission line which performs highly efficient photodetection in a simplified manner as compared to previous proposals. We extensively study the scattering properties of realistic wavepackets against this device, thereby computing the efficiency of the detector. We find that the detector has many operating modes, can detect detuned photons,  is robust against design imperfections and can be made broadband by using more than one absorbing element in the design.
\end{abstract}
\maketitle

\section{introduction}
The field of quantum circuits is undergoing a silent revolution, which started with the first superconducting qubits\cite{nakamura97,bouchiat98,mooij99,martinis02}, greatly advanced in the matter-wave interaction field\cite{blais04,wallraff04,chiorescu04}, and is now preparing the foundations of an entirely new technology: propagating quantum microwaves. The first ingredients in this new field are the generation of nonclassical propagating waves --either through qubits and cavities\cite{bozyigit10,eichler10}, or through nonlinearities\cite{mallet10}-- and the analysis of those fields, currently done using quantum homodyne detection techniques\cite{mariantoni05,mariantoni10,menzel10,bozyigit10,eichler10}. In order to consolidate and complete the field, we still lack two other ingredients: photon-photon interactions and single-shot photon detection and counting. In particular, photodetection is the ultimate and most desired goal. It is common to Quantum Optics and Quantum Information protocols, from trivial homodyne detection methods up to sophisticated all-optical quantum computing protocols\cite{knill01}. Developing such a tool in circuit-QED would open the door to quantum communication, quantum cryptography and general purpose quantum information processing with propagating photons.

\begin{figure}[t]
\includegraphics[width=0.95\columnwidth]{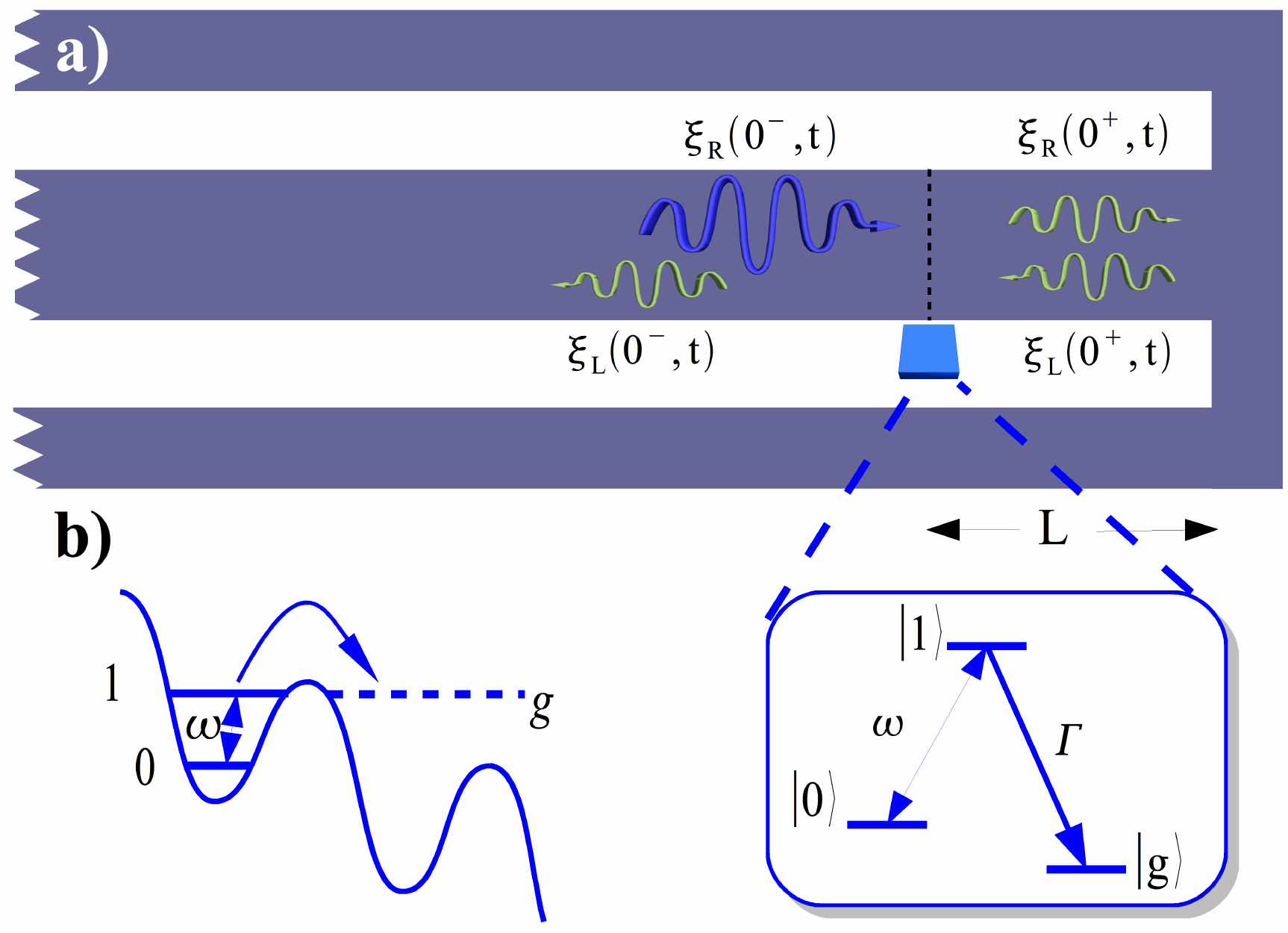}
\caption{Scheme of our microwave photon detector proposal. (a) The setup consists of a metastable quantum circuit positioned at a distance $L$ from the right mirror of a one-sided cavity forming a pseudo-cavity. (b) The quantum circuit can be made from a current-biased Josephson junction, in which a wash-board potential confines two metastable states that can decay into a continuum of current states.}
\label{Fig1}
\end{figure}

In a previous work\cite{romero09a,romero09b}, we identified photodetectors as the ultimate missing tool in circuit-QED, and helped in specifying the desired properties of such a device: it should be single-shot, work outside the cavity\cite{helmer09}, achieve great efficiency, be broadband and be passive. In that same work, we proposed a rather minimal design that performed the task\cite{romero09a,romero09b}: coupling phase-qubits to open transmission lines. In our design the phase qubit acts as a metastable three-level system which can absorb individual photons and transition into a third, easily detectable state, in a process that implements single-photon detection [Fig.~\ref{Fig1}b] with strict upper limit of 50\%. We showed that adding more qubits, this value could be easily increased up to 100\%. This also had the side effect of improving both the bandwidth and robustness of the detector.

In this work, we show that a slight modification of our design boosts its efficiency up to $100\%$ for a single qubit detector, without affecting the bandwidth or robustness of the original design. The small change consists of embedding the phase qubit in a semi-infinite line, at some distance from the end, which behaves as a perfect mirror. Qualitatively, in this new setup the end mirror allows incoming photons to bounce back from the end of the line and have several chances to be detected just by a single qubit. Alternatively, the setup can be seen as a one-dimensional implementation of the idea in Ref.~\onlinecite{pinotsi08}, by which a two-level system is made to absorb a photon whose wavefunction is the complex conjugate of that from an spontaneously emitted photon.

Our present work is also related to two recent developments. The first one is the implementation of a microwave photodetector using phase-biased Josephson junctions in Ref.~\onlinecite{chen10}. This setup contains some ingredients that are needed for the proposals in this and previous manuscripts~\cite{romero09a,romero09b}, and in particular its layout closely resembles the ones put forward in this manuscript. The second work is devoted to the study of the quasibound states that appear when a qubit is confronted with a mirror\cite{dong09}. Those resonances are to a large extent responsible for the high-efficiency and long interaction times between incoming photons and our detector. This is further evidenced in our study of photodetection when the photons are directly injected between the qubit and the mirror [Sect.~\ref{sec:right}].

This work is organized as follows. In Sect.~\ref{sec:full} we study in detail a design that consists of a three-level system sitting on a semi-infinite transmission line, and how it interacts with a finite-width propagating photon. We work in the strong interaction, rotating-wave approximation regime, in which the number of excitations is preserved, and develop an analytic approximation to the dynamics [Sect.~\ref{sec:dynamics}]. With these tools we can demonstrate that for a wide variety of parameters, a single photon may be perfectly absorbed by the three-level detector, even when it is detuned [Sect.~\ref{sec:result}]. In Sect.~\ref{sec:scattering} we develop a simplified theory based on scattering of plane waves which reproduces the previous results and allows us to study setups with more than one three-level system. The main result is that an increased number of absorbers enhances the robustness, the bandwidth and the overall performance of the detector. In Sect.~\ref{sec:right} we slightly modify our theory to study what happens when photons are not coming from the semi-infinite transmission line, but rather injected through the end of the line. We will show that efficient photodetection is still possible and mediated by quasi-localized states between the qubit and the mirror, at the expense of longer detection times. Finally in Sect.~\ref{sec:conclusions} we summarize our results.

\section{A qubit and a mirror}
\label{sec:full}

In this section, we discuss the simple setup of a semi-infinite transmission line coupled to a metastable quantum circuit located at a distance $L$ from the end of the waveguide, which acts as a mirror [See Fig.~\ref{Fig1}]. Studying the problem in real space, we derive the relevant equations describing the dynamics of the system. These models are used to study the scattering of a photon wavepacket showing that, under realistic conditions, it is completely absorbed by the metastable quantum circuit which represents the detector itself. More precisely, we demonstrate that when the photon is not reflected, the fraction that bounces back and forth between the absorber and the end mirror is also absorbed and thus detected. This will be the starting point for a more general and simpler theory in the following section.

\subsection{One absorber interacting with a single photon}
\label{sec:dynamics}

As sketched before, the basis of our work, as in our previous proposal~\cite{romero09a,romero09b}, is the real-space representation of a one-dimensional waveguide interacting with a single qubit\cite{shen05}. The model consists of a non-Hermitian Hamiltonian that contains terms for the metastable quantum circuits or ``absorbers'', modeled as three-level systems, and the radiation fields, $\psi_{R}$ and $\psi_{L},$ propagating to the right and to the left with group velocity $v_g$
\begin{eqnarray}\label{HamiltonianG}
H & = & \sum_{i}\hbar\left(\omega_{i}-i\frac{\Gamma_{i}}{2}\right)|1\rangle_{i}\langle1|  \\
 & + &i\hbar v_g \int\left[\psi_{L}^{\dagger}\partial_{x}\psi_{L}-\psi_{R}^{\dagger}\partial_{x}\psi_{R}\right]dx\nonumber \\
 & + & \sum_{i}\hbar V\int\delta(x-x_{i})\left[\left(\psi_{R}+\psi_{L}\right)|1\rangle_{i}\langle 0|+\mathrm{H.c.}\right]dx.\nonumber
\end{eqnarray}
Note how the interaction between photons and circuits is modeled using a delta potential of strength $V$ located at the positions of the latter, $x_i.$ In this notation, $|0\rangle$ and $|1\rangle$ represent the two states of the absorber connected by the photon [See Fig.~\ref{Fig1}b], $\Gamma_i$ stands for the decay rate from the metastable state $|1\rangle$, and $\omega_i$ is the frequency separation between $|0\rangle$ and $|1\rangle$.
 
The simplest scenario that we consider is a single photon interacting with one absorber placed at $x=0$, as shown in Fig.~\ref{Fig1}a. The photon coming from the left with energy $E=\hbar |k| v_g$, will exchange its excitation with the absorber such that the state of the system is\cite{shen05}  
\begin{equation}
|\Psi\rangle=\int\left[\xi_{R}(x)\psi_{R}^{\dagger}(x)+\xi_{L}(x)\psi_{L}^{\dagger}(x)\right]|0,vac\rangle+e(t)|1,vac\rangle.
\end{equation}
Here $\psi_{R,L}^\dagger(x)|0,vac\rangle$ is the state of a photon created at position $x$ and moving either to the right or to the left, while the absorber is in the metastable state $|0\rangle$. Also, $\xi_R(x,t)$ and $\xi_L(x,t)$ represent the wavefunction of a single photon moving to the right and to the left, respectively, while $|1,vac\rangle$ is a state with no photons and the absorber excited to unstable level $|1\rangle.$ Finally, $e(t)$ stands for the excited state population of the absorber. Solving the Schr\"odinger equation with the non-Hermitian Hamiltonian~(\ref{HamiltonianG}), leads to a set of equations containing the field and absorber amplitudes:
\begin{eqnarray}\label{syseq}
i\partial_{t}\xi_{R}(x,t) & = & -iv\partial_{x}\xi_{R}(x,t)+V\delta(x)e, \\
i\partial_{t}\xi_{L}(x,t) & = & +iv\partial_{x}\xi_{L}(x,t)+V\delta(x)e, \nonumber \\
\left(i\partial_{t}-\omega+i\frac{\Gamma}{2}\right)e & = & \frac{V}{2}\left[\xi_{R}^{+}+\xi_{R}^{-}+\xi_{L}^{+}+\xi_{L}^{-}\right],\nonumber
\end{eqnarray}
where we abbreviate $\xi_{R,L}^{\pm}(t):=\xi_{R,L}(0^{\pm},t).$ As explained elsewhere\cite{romero09b}, our quantum jump description allows us to compute the population of the level $\ket{g}$ as $P_g=1-||\Psi||^2$. Indeed, the value of $P_g$ at long times is what we call the detector efficiency and can be fully determined from the previous equations, after a few manipulations.

Note that two equations in~(\ref{syseq}) can be turned into boundary conditions around the absorber
\begin{equation}
\xi_{R}^{+} = \xi_{R}^{-}-i\frac{V}{v_g}e,\quad
\xi_{L}^{-} = \xi_{L}^{+}-i\frac{V}{v_g}e.
\end{equation}
This allows us to express the amplitude of the unstable state $|1\rangle$, in terms of the left and right incoming fields, that is
\begin{equation}
\left[i\partial_{t}-\omega+i\frac{\Gamma}{2}+i\frac{V^{2}}{v_g}\right]e=V\left[\xi_{R}^{-}+\xi_{L}^{+}\right].
\label{ampeEQ}
\end{equation}
The above procedure is standard in any single photon scattering problem, but in this case the mirror to the right end imposes another key boundary condition, which is a coupling between right and left propagating fields. More precisely, the only independent variable will be the field coming from the left, $\xi_{R}(0^{-},t)=\phi(t),$ since the incoming field from the right, $\xi_{L}(0^{+},t),$ is generated by the former, after being reflected by the mirror and affected by a phase factor $\kappa.$ In other words,
\begin{equation}
\xi_{L}^{+}(t)=\kappa\xi_{R}^{+}(t-a)=\kappa\phi(t-a)-\kappa i\frac{V}{v_g}e(t-a),
\end{equation}
with $a=2L/v_g$ depending on the distance between the absorber and the mirror, and the group velocity of the photons. This boundary condition provides us with a closed delay differential equation (DDE) for the amplitude of state $\ket{1}$
\begin{eqnarray}\label{ampeEQ1}
i\partial_{t}e(t) & = & \left[\omega-i\frac{\Gamma}{2}-i\frac{V^{2}}{v_g}\right]e(t)-\kappa i\frac{V^{2}}{v_g}e(t-a)\\
 & + & V\phi(t)+\kappa V\phi(t-a),\nonumber
\end{eqnarray}
thereby specifying the complete dynamics of the system for any incoming signal.

\subsection{Gaussian wavepacket}
\label{sec:result}

DDEs are very complicated mathematical objects which rarely have analytic solutions and which typically lead to nonlinear phenomena. In order to simplify the treatment, avoid critical behavior and get some understanding of the detection of realistic wavepackets, we will make some additional simplifications. More precisely, we will assume an incoming wavepacket with frequency $\omega_{0}$ and adiabatic modulation,
\begin{equation}
  \phi(t)=\chi(t)\exp(-i\omega_{0}t),\quad |\partial_t\chi|\ll \omega_0.
\end{equation}
This ansatz has various consquences for the dynamics. First of all, the absorber itself will evolve according to the main frequency, $e(t)=v_gx(t)\exp(-i\omega_{0}t)/V.$ Second, introducing the constants $\theta=\omega_{0}a$ and $a=v_g\Delta/V^{2},$ and making the change of variables $t=v\tau/V^{2}$ we will obtain a simplified equation
\begin{eqnarray}
i\partial_{\tau}x(\tau) & = & -i(1+\gamma)x(\tau)-izx(\tau-\Delta)\nonumber \\
& + & \chi(\tau)+z\chi(\tau-\Delta),
\label{diff_eq_x}
\end{eqnarray}
with only two free parameters
\begin{equation}
\gamma=\frac{v_g}{V^{2}}\left[\frac{\Gamma}{2}+i(\omega-\omega_{0})\right],\; z=\kappa e^{i\theta}.
\end{equation}
Finally, using the \textit{adiabatic approximation} that is the smoothness of the envelope, $|\partial_\tau\chi| \ll \omega_{0},$ we may replace $\chi(\tau-\Delta)$ by $\chi(\tau),$ and integrate the resulting equation
\begin{equation}
x(\tau)=-i(1+z)\int_{-\infty}^{\tau}e^{-(1+\gamma+z)(\tau-s)}\chi(s)ds.
\label{xtau}
\end{equation}
The whole problem has simplified considerably. One may now study, for instance, a normalized Gaussian wavepacket
\begin{equation}\label{Gausswave}
\chi(\tau)=\frac{1}{\sqrt{\sigma\sqrt{\pi}}}
\exp\left[-\tau^2/(2\sigma^{2})\right],
\end{equation}
and how it is scattered by the three-level system. The Gaussian form is chosen for convenience, but it is in no way essential for the results. This Gaussian has the advantage that in the limit $\sigma \rightarrow \infty,$ it contains the case of infinite plane waves, a limit which we used in previous works and which we would like to recover. However, as long as the wavepacket remains adiabatic, that is $\sigma\gg\omega_0^{-1},$ none of the results will depend dramatically on its precise shape, as we confirmed numerically.

\begin{figure}[t]\flushleft
\includegraphics[width=\columnwidth]{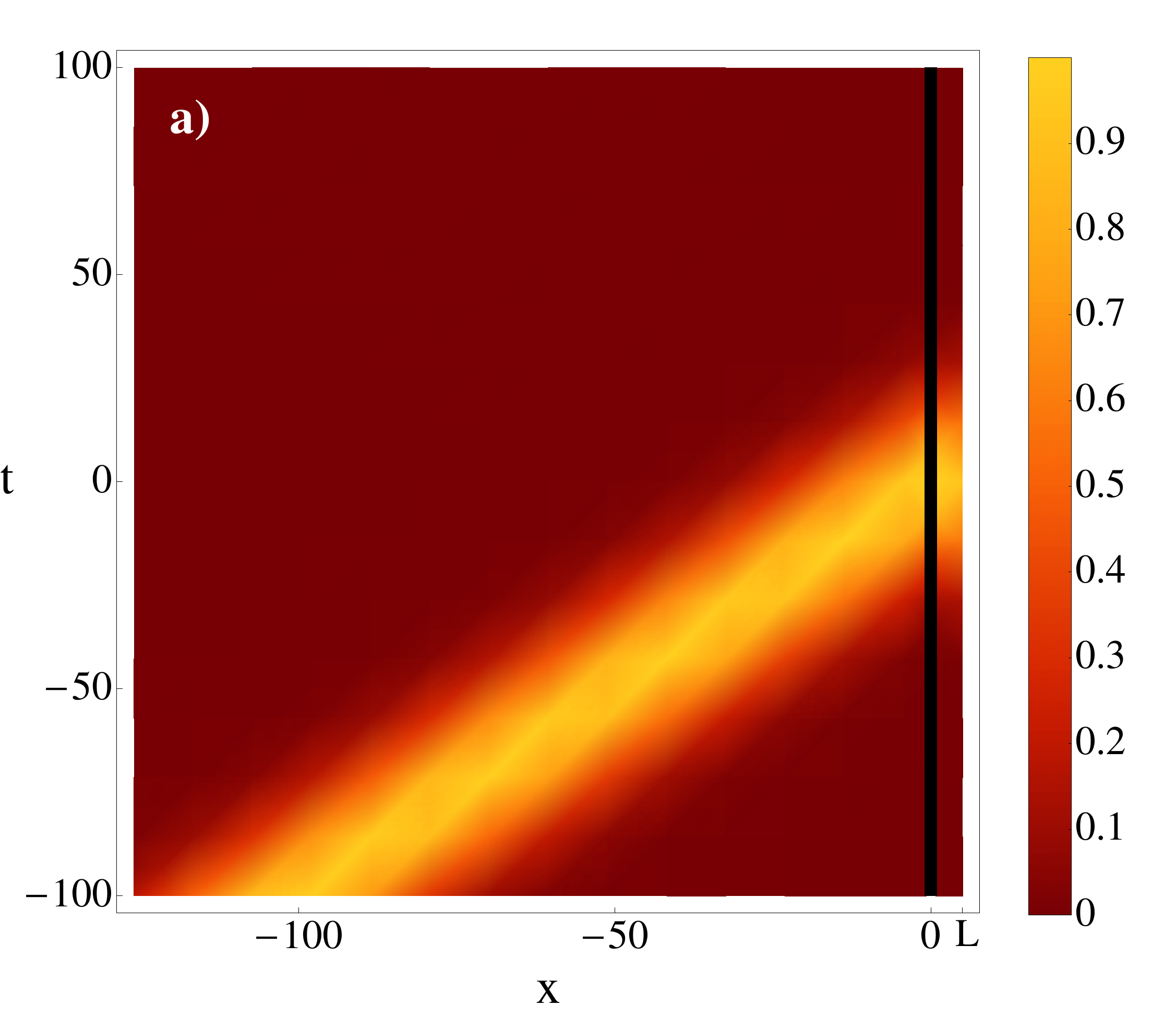}
\newline
\includegraphics[width=0.95\columnwidth]{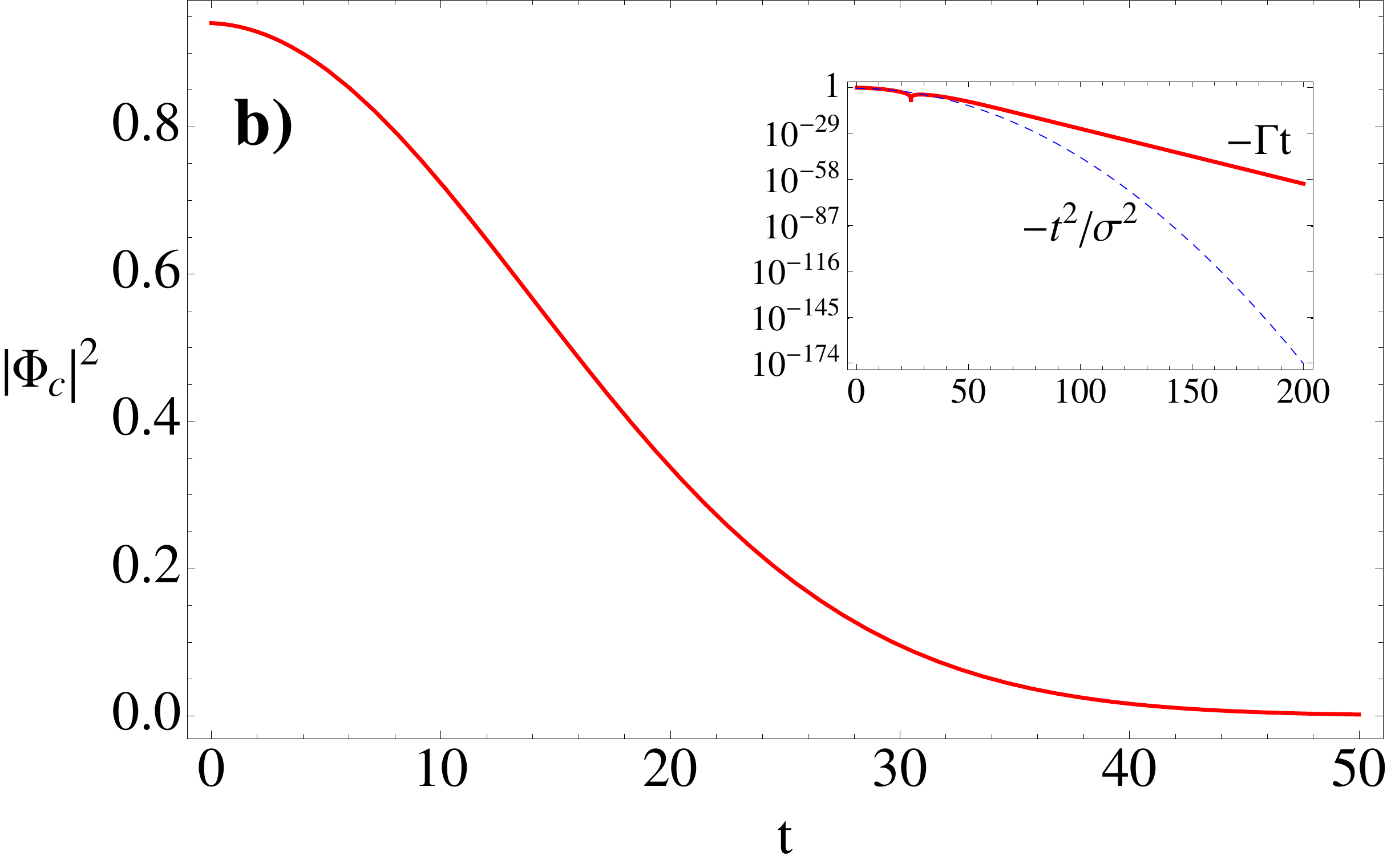}
 \caption{a) Time evolution of an incident wavepacket that undergoes
    no reflection leading to a confined field to the right of the absorber. b) Decay of the field
    inside the absorber-mirror cavity. The inset shows a logarithmic plot of the field amplitude exhibiting a time scale with decay.}
 \label{noreflection}
\end{figure}
We are now in a position to compute the transmitted and reflected wave packets, $\xi_R^{+}(t), \xi_L^{-}(t),$ the dynamics of the detector, $x(t),$ and even the detection probability mentioned before, all as a function of the parameters $\sigma,\gamma$ and $z.$ The first result that we show in Fig.~\ref{noreflection}a is that there indeed exist configurations for which no incoming photon is reflected. In such setups the photon tunnels through the qubit and bounces back and forth between the qubit and the mirror. If this process continued indefinitely, our system could not perform as a photodetector, as it would never capture the photon and switch. In order for the photodetection to succeed, the three-level system must be able to absorb completely the confined field and undergo an irreversible transition to the ``click'' state $|g\rangle.$ Fortunately, as Fig.~\ref{noreflection}b shows, the population of the field inside the qubit-mirror pseudo-cavity dissipates very quickly, and in a timescale determined by the decay channel of the absorber, $\Gamma^{-1},$ the absorber fully detects the confined photon.

\begin{figure}[t]
 \includegraphics[width=0.85\columnwidth]{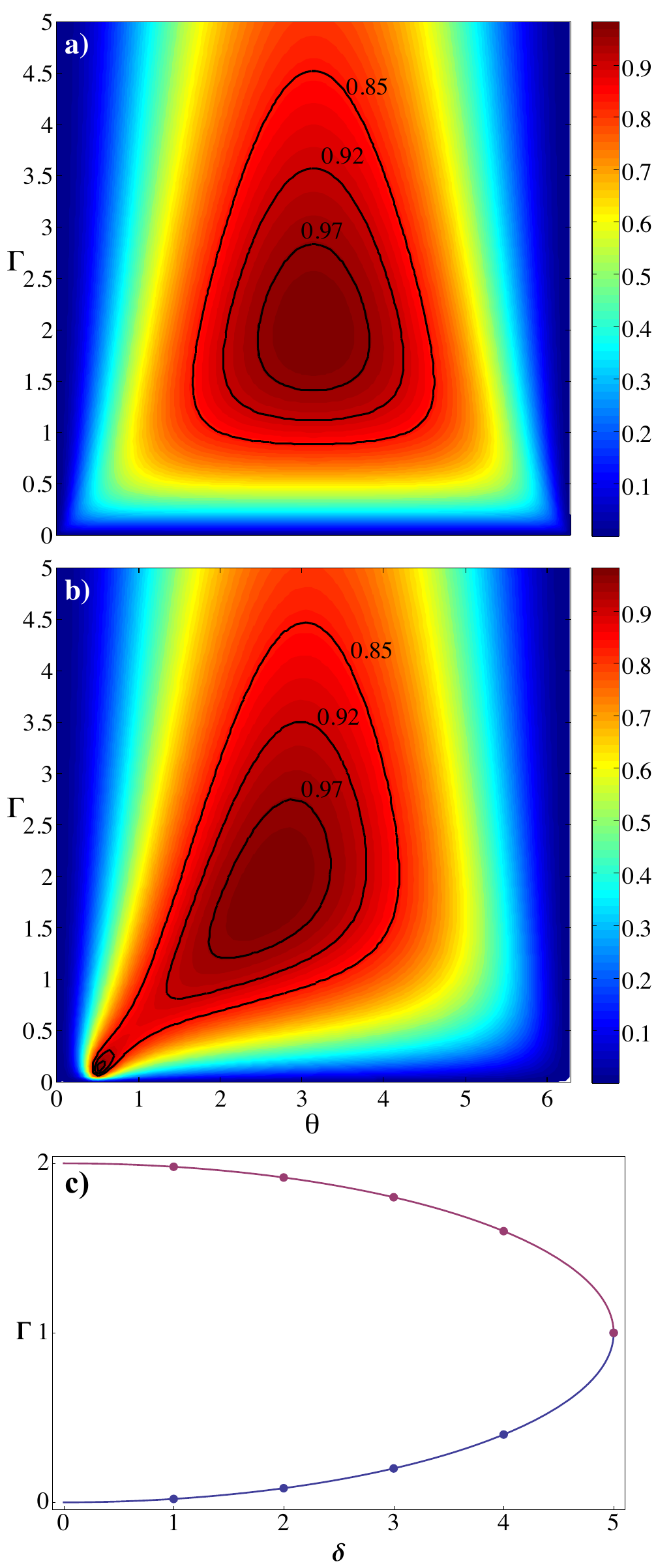}
%\newline\includegraphics[width=0.9\columnwidth]{abs_eff_det05}
%\newline \includegraphics[width=0.9\columnwidth]{zeros}
 \caption{a) Detection efficiency as a function of the decay $\gamma$
   and the phase $\theta$ when the three level system is on resonance
   with the incident photon. b) Real part of the detection efficiency for the
   off-resonance case. c) Position of the efficiency maxima
   as a function of the detuning (off-resonance case).}
 \label{Reflection}
\end{figure}

The above results give us a hint that a single absorber with a mirror could reach $100\%$ detection efficiency, but this result must be confirmed for a larger variety of experimental parameters. In order to make the study more systematic, we introduce the detector efficiency as the fraction of the wavepacket that was absorbed, given by
\begin{equation}
\alpha=1-\frac{\int_{-\infty}^{\infty}|\xi_L(0^{-},t)|^2}{\int_{-\infty}^{\infty}|\xi_R(0^{-},t)|^2}.
\label{abs_eff}
\end{equation}
This value is computed numerically for different photon profiles, $\sigma,$ and varying setup parameters, $\omega, \omega_0, \theta$ and $\Gamma.$ With respect to the pulse width, we have found that any value of $\sigma > 10$ gives approximately the same result. For the other parameters we have to distinguish the resonant and nonresonant cases, and in the latter study the dependence of the efficiency on the detuning, $\delta=\omega-\omega_0.$ As shown in Fig.~\ref{Reflection}a, for a resonant incident photon the efficiency reaches a maximum of $100\%$ around $\theta=\pi,\;\Gamma=2$~(Fig. \ref{Reflection}a), where $\Gamma$ is in units of $V^2/v_g$. When the photon is off-resonant, $\delta\neq 0,$ we obtain two remarkable results. First of all, the perfect photodetection is still possible, and second, this happens for two different sets of parameters, as shown in Fig.~\ref{Reflection}b. The relative position of the two maxima depends on the coupling strength $V.$ These solutions aproach each other [Fig.~\ref{Reflection}c] until the detuning reaches a threshold $\delta\leq V^2/v_g,$ where the two solutions merge and disappear. Using parameters in the range of Ref.~\onlinecite{romero09a} this sets the limit of the bandwidth around $\delta \sim 10-100\textrm{MHz}$ for just a single detector, but it increases for larger couplings.

\section{Scattering theory}
\label{sec:scattering}

\begin{figure}[t]\flushleft
\includegraphics[width=\columnwidth]{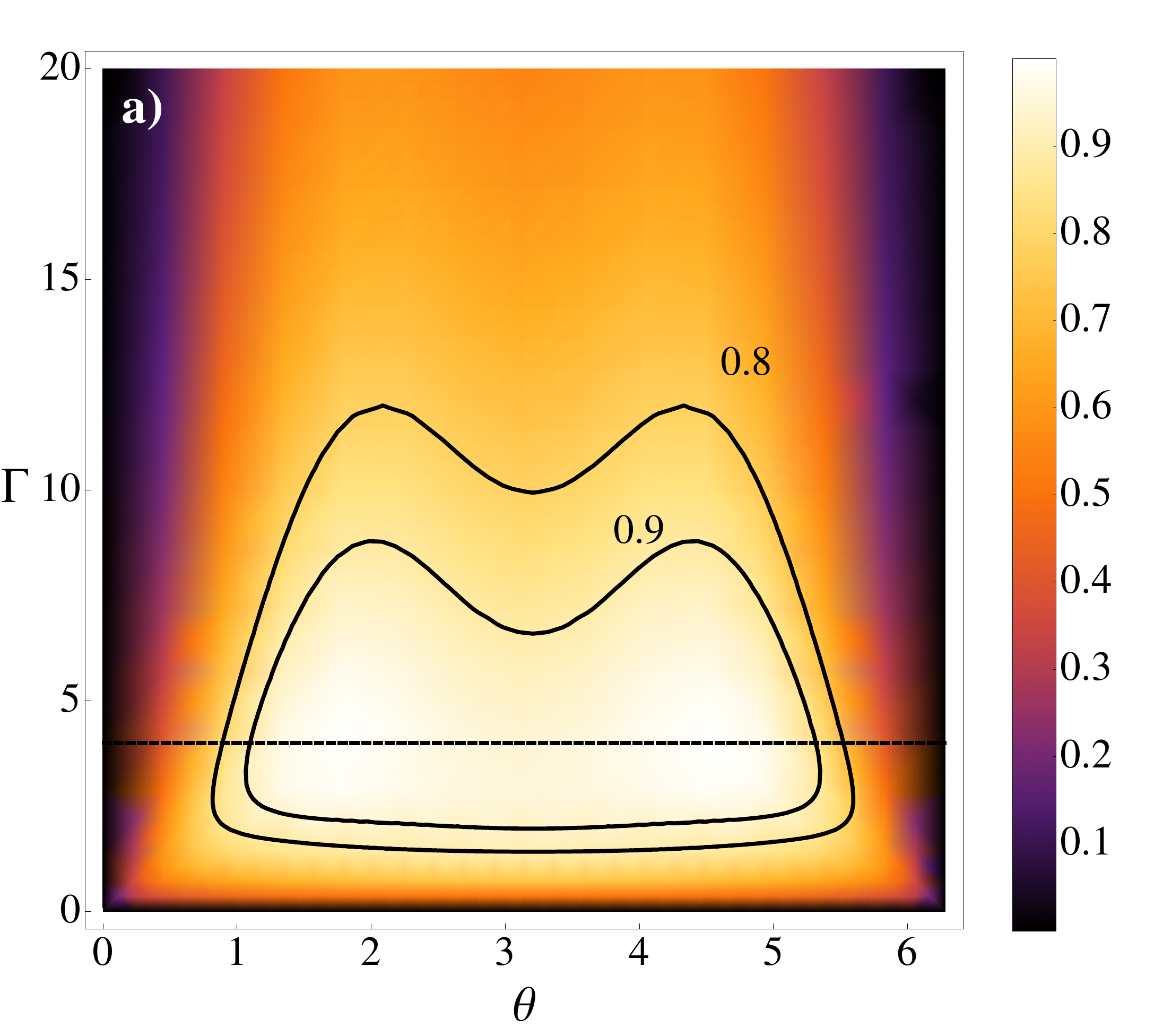}
\includegraphics[width=0.9\columnwidth]{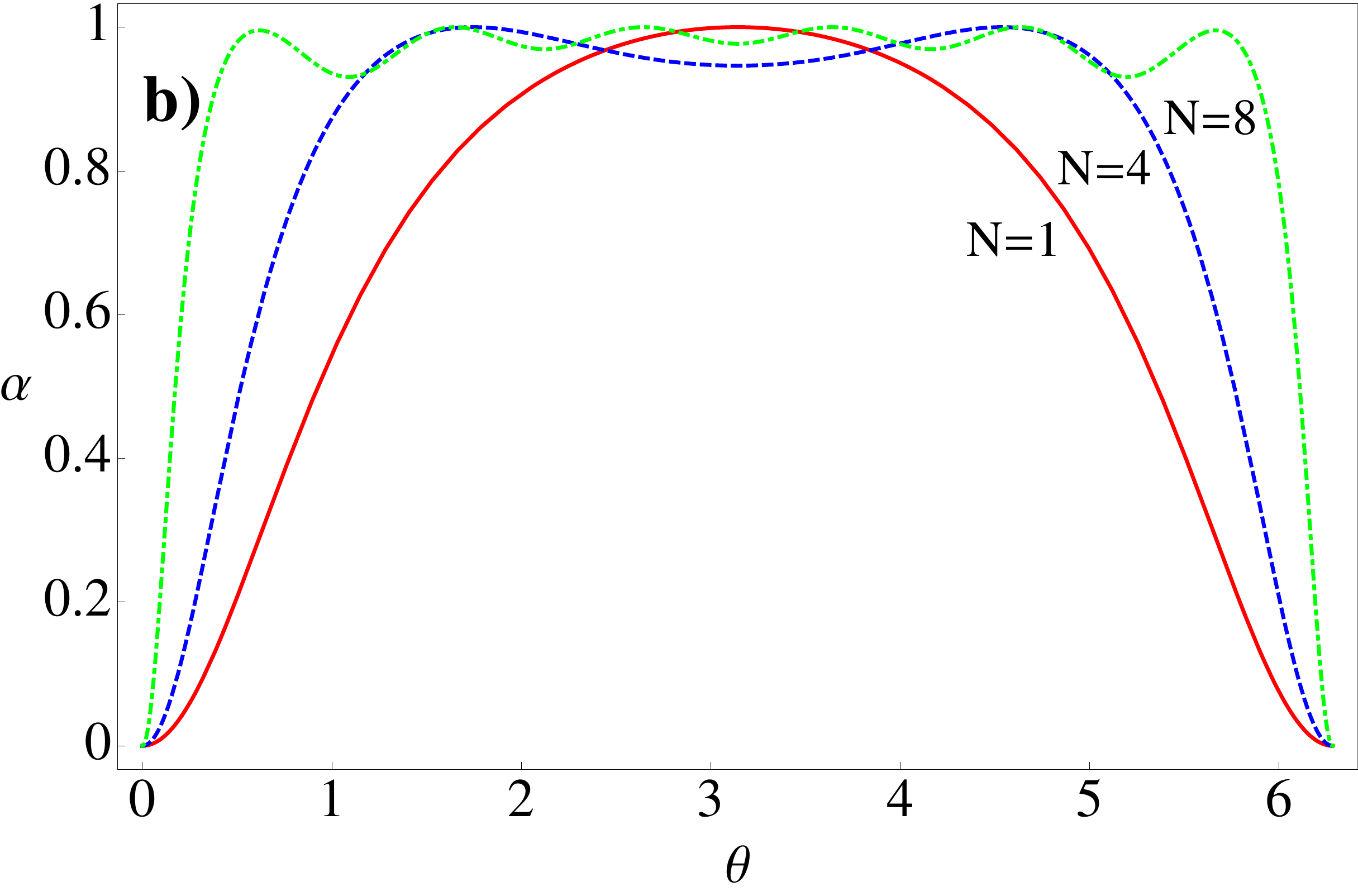}
\includegraphics[width=0.9\columnwidth]{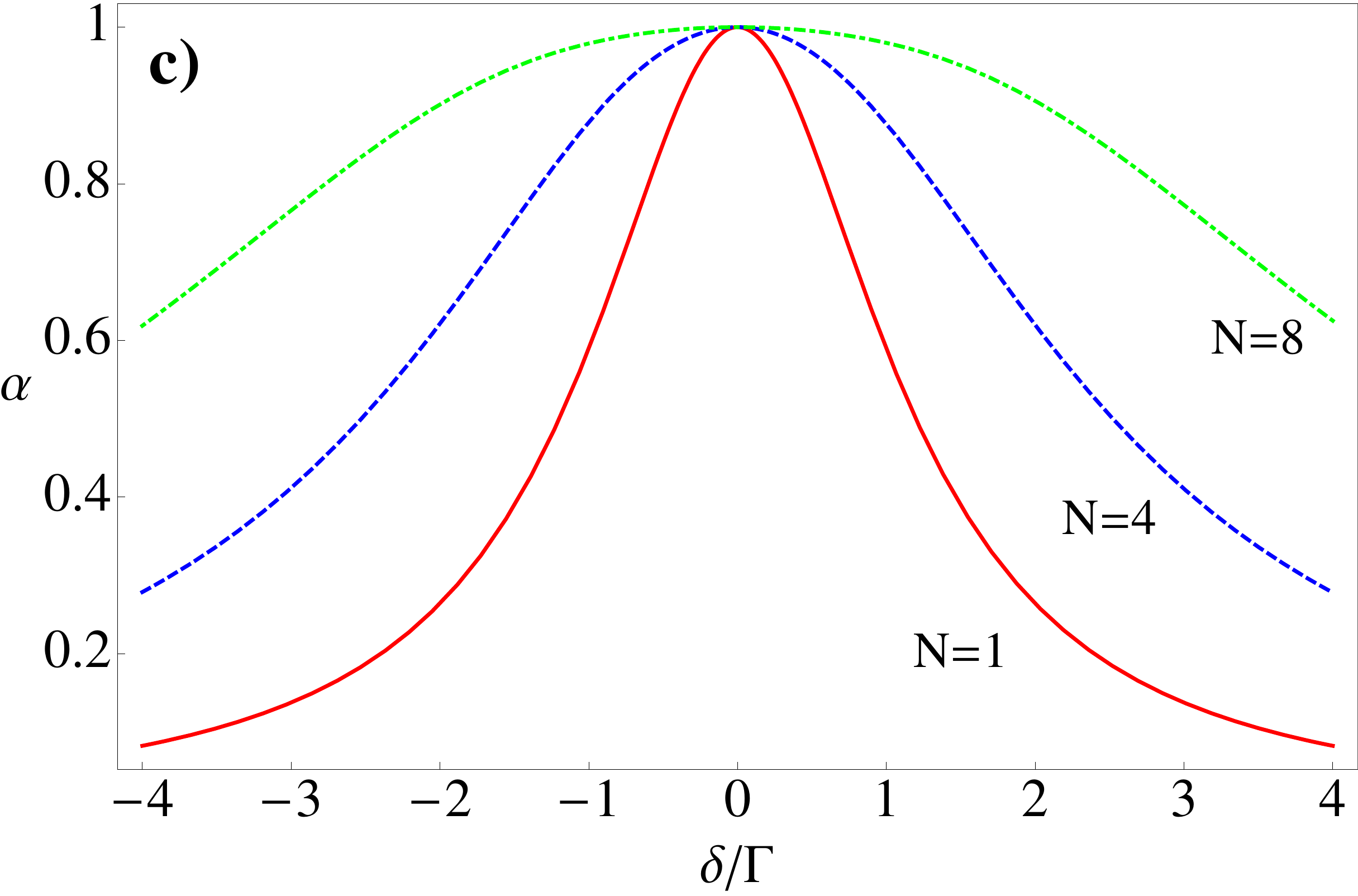}
\caption{a)Absorption efficiency for 4 qubits as a function of the
 decay and phase $\theta$. b) Performing a cut along the optimal value of $\Gamma$ (dashed
 line), we compute the efficiency as a function of $\theta$ for
 $N=1,4$ and $8$ absorbers. c) Efficiency dependence with detuning rate}
\label{more_absorbers}
\end{figure}

In the previous section we have demonstrated two important results. The first one is that the scattering of a realistic wavepacket through a single three-level system indicates the existence of a regime of perfect photodetection. The second one is that we can analytically compute all scattering properties for a sufficienctly large wavepacket and that these values are almost insensitive to the wavepacket size. This result motivates us to replace the previous formalism with a simpler one, based on the scattering of plane-waves through one or multiple three-level systems. This method, developed in Ref.~\onlinecite{shen05} and applied in our photodetector works\cite{romero09a,romero09b}, has the advantage that it scales well to setups with multiple detectors, an ingredient which is crucial for enhancing the robustness and the bandwidth of the detector.

Consider an incident monochromatic beam interacting with more qubits, using the scattering theory developed in \cite{shen05,romero09a,romero09b}. The idea is that the fields on the left and on the right of the absorbers are related by a scattering matrix
\begin{equation}
\left(\begin{array}{c}
\xi_R'\\
\xi_L'\end{array}\right)=T\left(\begin{array}{c}
\xi_R\\
\xi_L\end{array}\right),
\end{equation}
where $T$ stands for the transfer matrix, and takes the form
\begin{equation}
T=\prod_j^N e^{i\frac{2\pi L_j}{\lambda}\sigma_z}T_j, \quad 
T_j=\left(\begin{array}{cc}
1-1/\gamma & -1/\gamma\\
1/\gamma & 1+1/\gamma\end{array}\right)\label{Tmatrix}
\end{equation}
Compared to Ref.\onlinecite{romero09a}, the main difference now is that after leaving the scatterers and confronting the mirror, the field has to satisfy a boundary condition
\begin{equation}
\left(\begin{array}{c}
1\\
\kappa\end{array}\right)=\exp(i\theta\sigma^{z})T\left(\begin{array}{c}
\xi_R\\
\xi_L\end{array}\right).
\end{equation}
The parameter $\theta$ is the phase acquired by the photon between the last scatterer and the mirror, while $\kappa$ is the boundary condition for the mirror to have zero field, typically $-1.$ The previous equations hide a relation between the incoming field $\xi_R^{-}$ and the reflected one $\xi_L^{-},$ which can be revealed by projecting onto an orthogonal subspace
\begin{equation}
\left(\begin{array}{cc}
\kappa & -1\end{array}\right)\exp(i\theta\sigma^{z})T\left(\begin{array}{c}
\xi_R\\
\xi_L\end{array}\right)=0.
\end{equation}
In the case of a single absorber ($N=1$), we directly obtain an analytic expression for the outgoing field,
\begin{equation}
\xi_{L}= \left[\frac{\gamma(z+1)}{1+\gamma+z}-1\right]\xi_R,
\end{equation}
which becomes exactly zero for
\begin{equation}
  \label{eq:1}
  \gamma=1+z^{-1},
\end{equation}
reflecting the limit in which no photon is reflected and all photons are absorbed, in perfect agreement with the exact results for gaussian wavepackets developed in the previous section.

Using this formalism, we can go beyond one absorber, studying the optical properties of a setup with multiple three-level systems in front of a mirror. Inspired by our previous works we expect that a setup with multiple scatterers will help, first, by increasing the robustness of the detector and, second, by enlarging the band of frequencies for which perfect detection takes place. Furthermore, as shown in those works, the way in which we place the absorbers is very relevant, as placing them too close together does not have any influence in the detector efficiency or bandwidth. For simplicity, we will adopt the optimal configuration from the open line, with equally spaced absorbers. From the elements of the transfer matrix given by Eq.~(\ref{Tmatrix}), we can compute the absorption efficiency~(\ref{abs_eff}) using the formula
\begin{equation}
\alpha=1-\left\vert\frac{T_{11}+e^{i\theta}T_{12}}{T_{21}+e^{i\theta}T_{22}}\right\vert^2,
\end{equation}
where now $T_{ij}$, depends on the number of absorbers, $N,$ and the previous two parameters, $\Gamma, \theta.$

As an illustration, in Fig.~\ref{more_absorbers} we show three plots that demonstrate the enhanced bandwidth and decreased sensitivity to the qubit and setup properties, $\Gamma, \theta.$ To start with, let us look at Fig.~\ref{more_absorbers}a, which plots the detector efficiency for $N=4$ absorbers. Compared with Fig.~\ref{Reflection}a, the maximum efficiency is extended to a larger region of mirror separations, now centered around $\pi/2,3\pi/2,$ and tolerates also a larger set of decay rates, $\Gamma.$ This is further confirmed when we study the evolution of the efficiency for increasing number of absorbers. For instance, Fig.~\ref{more_absorbers}b represents the efficiency as a function of the phase $\theta=4\pi L/\lambda$, where $L$ stands for the distance between absorbers. Notice that, for $N=8$ absorbers, the efficiency reaches more than 90\% almost independently of $\theta,$ becoming less important the relative position between absorbers. A similar effect happens with the detuning, and as Fig.~\ref{more_absorbers}c shows, the set of multiple detectors very quickly  acquires a large bandwidth, even faster than in our previous works\cite{romero09a,romero09b}. Finally, while we do not explicitly show results here, we expect that this setup will have the same robustness against imperfection as was demonstrated for our original setup\cite{romero09b}.

\section{Detecting through the mirror}
\label{sec:right}

\begin{figure}[t]\flushleft
\includegraphics[width=\columnwidth]{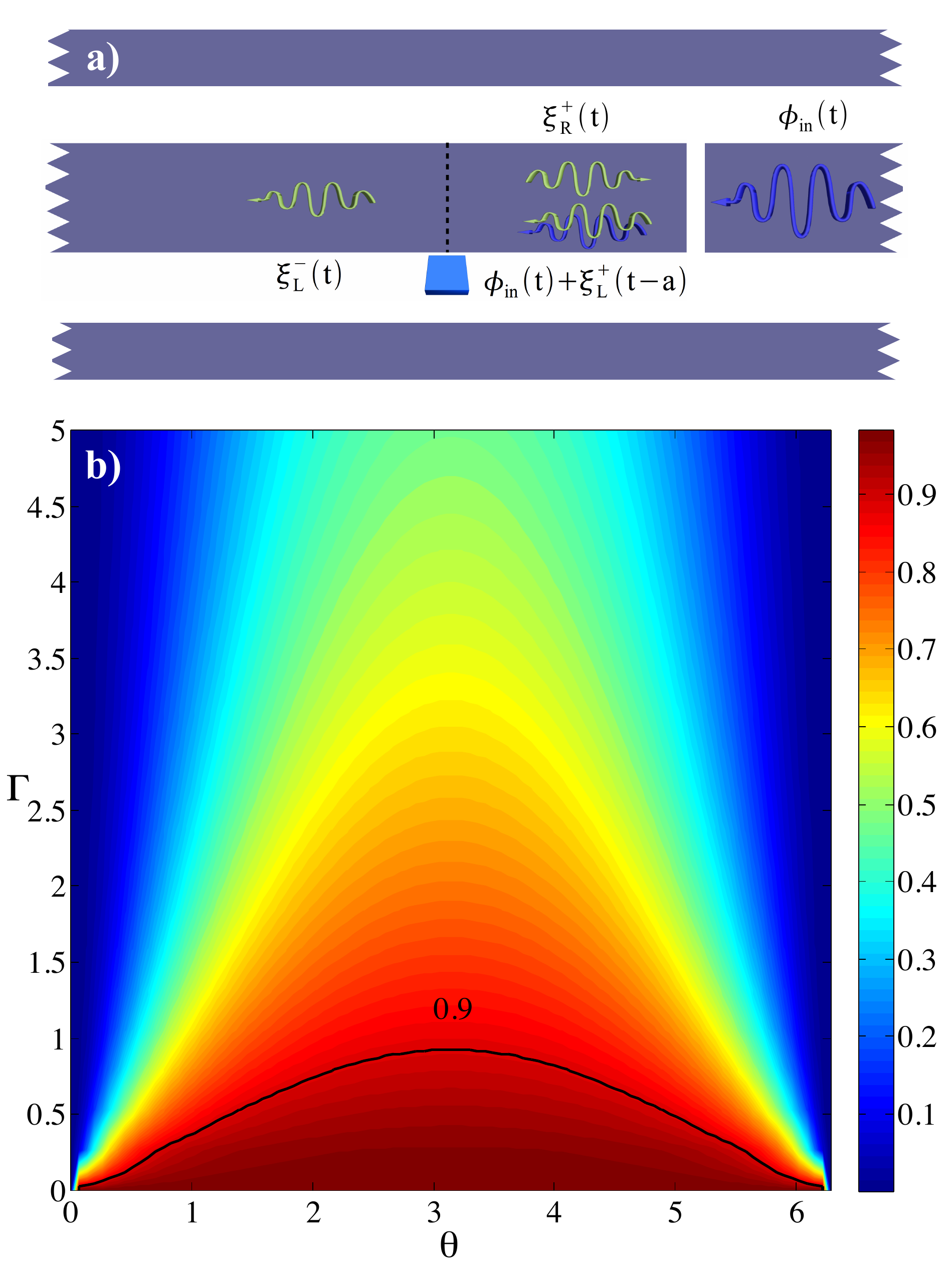}
\caption{a) Photodetector scheme for an incident photon coming from the right. b) Detection efficiency associated to this detector.}
\label{exp}
\end{figure}
On looking at our setup it arises a natural question: what happens if the photon is not coming from the semi-infinite transmission line, but instead it ``tunnels'' through the mirror, which is not perfect. This is an interesting question for a number of reasons. The first one is that if the photon is directly injected between the mirror and the cavity it has a great chance to probe quasi-bound states existing between both, providing further evidence that the qubit and the mirror form a pseudo-cavity\cite{dong09}. The second reason is that this setup is close to the recent experiment\cite{chen10}  which demonstrates the photodetection capabilities of a phase-biased junction.

Describing a semi-infinite line with an imperfect mirror would severely depart from the methods introduced in this paper, requiring the introduction of environments, decoherence and master equations. Fortunately, there is a simple ``toy'' model that contains the essential ingredients of the problem and which can still be treated with the scattering formalism. In our model the photon is directly tunneling in between the qubit and the mirror, as shown in Fig.\ref{exp}a, and the only way it may leak is passing through the qubit again. The incoming photon profile will be denoted by $\phi_{in}(t)$ and we expect tha the wavepacket is partially trapped into the pseudocavity formed by the mirror and the absorber, and is partially transmitted. Working with our previous single-photon formalism we obtain the following set of equations for the field amplitudes:
\begin{eqnarray}
\xi_{R}^{-} (t) &=& 0, \quad
\xi_{R}^{+}(t) =\xi_{R}^{-}(t) -i\frac{V}{v_g}e (t) ,\\
\xi_{L}^{-}(t)  &=& \xi_{L}^{+}(t) -i\frac{V}{v_g}e (t) ,\quad
\xi_{L}^{+}(t) = \phi_{in}(t)+\kappa\xi_{R}^{+}(t-a), \nonumber
\end{eqnarray}
where the presence of the time-delayed amplitude field $\xi_R^{+}$ is due to the iterative feedback with the mirror. Using the same tools, we can now compute the detection efficiency associated to the switching process of the three level system. As shown in Fig.\ref{exp}b, the contour plots of the efficiency suffer a radical change. While the maxima are still at a resonant distance between mirror and qubit, the $100\%$ detection efficiency is strictly achieved for $\Gamma\rightarrow 0.$

The previous analytical results have a very clear interpretation. In order to have a large detection efficiency, the photon has to spend a long time bouncing between the mirror and the qubit. However, as shown previously\cite{shen05}, the qubit only acts as a perfect mirror strictly for $\Gamma=0.$ The consequence is that in this setup the decay time of the three level system, $\Gamma,$ must approach zero to increase the efficiency, and at the same time the detection time diverges as $1/\Gamma.$ In other words, while this setup seems quantitatively similar to the previous ones, it does not work in practice, because first the tunneling probability of the photon through the mirror will be small and second the detection times are so long that the process will be damaged by decoherence and losses.

\section{Summary and conclusions}
\label{sec:conclusions}

The main result of this work is that a three-level system, implemented as a phase qubit, a biased Josephson junction, or whatever it seems adequate, is a perfect photodetector. The efficiency of this device is only limited by how it couples to the waves that contain the photons it has to detect. In particular, it seems that the setup we originally proposed\cite{romero09a,romero09b} can be improved by replacing a completely open transmission line with a semi-infinite line that brings the photons to the detector, allowing repeated interactions. We have studied this absorber-mirror system in detail, demonstrating that when an incoming photon passes through the qubit without reflection it is actually absorbed, and detected. This first result is quite important, because it rules out that the photon gets trapped in a metastable confined state between the absorber and the mirror\cite{chen10}, and because it allows us to develop a much simplified theory based on the scattering of plane waves. With this theory we confirm the $100\%$ efficiency of a single absorber, and extend our design to include multiple qubits in front of a mirror, a setup that shows enhanced bandwidth and very much decreased sensitivity to the detector properties.

\section*{Acknowledgements}

We thank Anders S{\o}rensen for pointing out the relation between our work and Refs.\onlinecite{pinotsi08,witthaut10}, and the discussions that sparked this research. This work was supported by Spanish MICINN Project  FIS2009-10061, and CAM research consortium QUITEMAD S2009-ESP-1594. B.P. acknowledges funding by CSIC JAE-PREDOC2009 Grant.  G.R acknowledges funding from Juan de la Cierva program.  G.J and C.M.W. thank funding from Swedish Research Council (VR) and European Research Council. E.S acknowledges funding from Basque Government Grant IT472-10, Spanish MICINN project FIS2009-12773-C02-01, and SOLID European project.

%\bibliographystyle{apsrev}
%\bibliography{cqed}

\begin{thebibliography}{22}
\expandafter\ifx\csname natexlab\endcsname\relax\def\natexlab#1{#1}\fi
\expandafter\ifx\csname bibnamefont\endcsname\relax
  \def\bibnamefont#1{#1}\fi
\expandafter\ifx\csname bibfnamefont\endcsname\relax
  \def\bibfnamefont#1{#1}\fi
\expandafter\ifx\csname citenamefont\endcsname\relax
  \def\citenamefont#1{#1}\fi
\expandafter\ifx\csname url\endcsname\relax
  \def\url#1{\texttt{#1}}\fi
\expandafter\ifx\csname urlprefix\endcsname\relax\def\urlprefix{URL }\fi
\providecommand{\bibinfo}[2]{#2}
\providecommand{\eprint}[2][]{\url{#2}}

\bibitem[{\citenamefont{Nakamura et~al.}(1997)\citenamefont{Nakamura, Chen, and
  Tsai}}]{nakamura97}
\bibinfo{author}{\bibfnamefont{Y.}~\bibnamefont{Nakamura}},
  \bibinfo{author}{\bibfnamefont{C.~D.} \bibnamefont{Chen}}, \bibnamefont{and}
  \bibinfo{author}{\bibfnamefont{J.~S.} \bibnamefont{Tsai}},
  \bibinfo{journal}{Phys. Rev. Lett.} \textbf{\bibinfo{volume}{79}},
  \bibinfo{pages}{2328} (\bibinfo{year}{1997}).

\bibitem[{\citenamefont{Bouchiat et~al.}(1998)\citenamefont{Bouchiat, Vion,
  Joyez, Esteve, and Devoret}}]{bouchiat98}
\bibinfo{author}{\bibfnamefont{V.}~\bibnamefont{Bouchiat}},
  \bibinfo{author}{\bibfnamefont{D.}~\bibnamefont{Vion}},
  \bibinfo{author}{\bibfnamefont{P.}~\bibnamefont{Joyez}},
  \bibinfo{author}{\bibfnamefont{D.}~\bibnamefont{Esteve}}, \bibnamefont{and}
  \bibinfo{author}{\bibfnamefont{M.~H.} \bibnamefont{Devoret}},
  \bibinfo{journal}{Phys. Scr} \textbf{\bibinfo{volume}{T76}},
  \bibinfo{pages}{165} (\bibinfo{year}{1998}).

\bibitem[{\citenamefont{Mooij et~al.}(1999)\citenamefont{Mooij, Orlando,
  Levitov, Tian, van~der Wal, and Lloyd}}]{mooij99}
\bibinfo{author}{\bibfnamefont{J.~E.} \bibnamefont{Mooij}},
  \bibinfo{author}{\bibfnamefont{T.~P.} \bibnamefont{Orlando}},
  \bibinfo{author}{\bibfnamefont{L.}~\bibnamefont{Levitov}},
  \bibinfo{author}{\bibfnamefont{L.}~\bibnamefont{Tian}},
  \bibinfo{author}{\bibfnamefont{C.~H.} \bibnamefont{van~der Wal}},
  \bibnamefont{and} \bibinfo{author}{\bibfnamefont{S.}~\bibnamefont{Lloyd}},
  \bibinfo{journal}{Science} \textbf{\bibinfo{volume}{285}},
  \bibinfo{pages}{1036} (\bibinfo{year}{1999}).

\bibitem[{\citenamefont{Martinis et~al.}(2002)\citenamefont{Martinis, Nam,
  Aumentado, and Urbina}}]{martinis02}
\bibinfo{author}{\bibfnamefont{J.~M.} \bibnamefont{Martinis}},
  \bibinfo{author}{\bibfnamefont{S.}~\bibnamefont{Nam}},
  \bibinfo{author}{\bibfnamefont{J.}~\bibnamefont{Aumentado}},
  \bibnamefont{and} \bibinfo{author}{\bibfnamefont{C.}~\bibnamefont{Urbina}},
  \bibinfo{journal}{Phys. Rev. Lett.} \textbf{\bibinfo{volume}{89}},
  \bibinfo{pages}{117901} (\bibinfo{year}{2002}).

\bibitem[{\citenamefont{Blais et~al.}(2004)\citenamefont{Blais, Huang,
  Wallraff, Girvin, and Schoelkopf}}]{blais04}
\bibinfo{author}{\bibfnamefont{A.}~\bibnamefont{Blais}},
  \bibinfo{author}{\bibfnamefont{R.-S.} \bibnamefont{Huang}},
  \bibinfo{author}{\bibfnamefont{A.}~\bibnamefont{Wallraff}},
  \bibinfo{author}{\bibfnamefont{S.~M.} \bibnamefont{Girvin}},
  \bibnamefont{and} \bibinfo{author}{\bibfnamefont{R.~J.}
  \bibnamefont{Schoelkopf}}, \bibinfo{journal}{Phys. Rev. A}
  \textbf{\bibinfo{volume}{69}}, \bibinfo{pages}{062320}
  (\bibinfo{year}{2004}).

\bibitem[{\citenamefont{Wallraff et~al.}(2004)\citenamefont{Wallraff, Schuster,
  Blais, Frunzio, Huang, Majer, Kumar, Girvin, and Schoelkopf}}]{wallraff04}
\bibinfo{author}{\bibfnamefont{A.}~\bibnamefont{Wallraff}},
  \bibinfo{author}{\bibfnamefont{D.}~\bibnamefont{Schuster}},
  \bibinfo{author}{\bibfnamefont{A.}~\bibnamefont{Blais}},
  \bibinfo{author}{\bibfnamefont{L.}~\bibnamefont{Frunzio}},
  \bibinfo{author}{\bibfnamefont{R.-S.} \bibnamefont{Huang}},
  \bibinfo{author}{\bibfnamefont{J.}~\bibnamefont{Majer}},
  \bibinfo{author}{\bibfnamefont{S.}~\bibnamefont{Kumar}},
  \bibinfo{author}{\bibfnamefont{S.~M.} \bibnamefont{Girvin}},
  \bibnamefont{and} \bibinfo{author}{\bibfnamefont{R.~J.}
  \bibnamefont{Schoelkopf}}, \bibinfo{journal}{Nature}
  \textbf{\bibinfo{volume}{431}}, \bibinfo{pages}{162} (\bibinfo{year}{2004}).

\bibitem[{\citenamefont{{Chiorescu} et~al.}(2004)\citenamefont{{Chiorescu},
  {Bertet}, {Semba}, {Nakamura}, {Harmans}, and {Mooij}}}]{chiorescu04}
\bibinfo{author}{\bibfnamefont{I.}~\bibnamefont{{Chiorescu}}},
  \bibinfo{author}{\bibfnamefont{P.}~\bibnamefont{{Bertet}}},
  \bibinfo{author}{\bibfnamefont{K.}~\bibnamefont{{Semba}}},
  \bibinfo{author}{\bibfnamefont{Y.}~\bibnamefont{{Nakamura}}},
  \bibinfo{author}{\bibfnamefont{C.~J.~P.~M.} \bibnamefont{{Harmans}}},
  \bibnamefont{and} \bibinfo{author}{\bibfnamefont{J.~E.}
  \bibnamefont{{Mooij}}}, \bibinfo{journal}{\nat}
  \textbf{\bibinfo{volume}{431}}, \bibinfo{pages}{159} (\bibinfo{year}{2004}).

\bibitem[{\citenamefont{{Bozyigit} et~al.}(2010)\citenamefont{{Bozyigit},
  {Lang}, {Steffen}, {Fink}, {Baur}, {Bianchetti}, {Leek}, {Filipp}, {da
  Silva}, {Blais} et~al.}}]{bozyigit10}
\bibinfo{author}{\bibfnamefont{D.}~\bibnamefont{{Bozyigit}}},
  \bibinfo{author}{\bibfnamefont{C.}~\bibnamefont{{Lang}}},
  \bibinfo{author}{\bibfnamefont{L.}~\bibnamefont{{Steffen}}},
  \bibinfo{author}{\bibfnamefont{J.~M.} \bibnamefont{{Fink}}},
  \bibinfo{author}{\bibfnamefont{M.}~\bibnamefont{{Baur}}},
  \bibinfo{author}{\bibfnamefont{R.}~\bibnamefont{{Bianchetti}}},
  \bibinfo{author}{\bibfnamefont{P.~J.} \bibnamefont{{Leek}}},
  \bibinfo{author}{\bibfnamefont{S.}~\bibnamefont{{Filipp}}},
  \bibinfo{author}{\bibfnamefont{M.~P.} \bibnamefont{{da Silva}}},
  \bibinfo{author}{\bibfnamefont{A.}~\bibnamefont{{Blais}}},
  \bibnamefont{et~al.}, \bibinfo{journal}{ArXiv e-prints}
  (\bibinfo{year}{2010}), \eprint{1002.3738}.

\bibitem[{\citenamefont{{Eichler} et~al.}(2010)\citenamefont{{Eichler},
  {Bozyigit}, {Lang}, {Steffen}, {Fink}, and {Wallraff}}}]{eichler10}
\bibinfo{author}{\bibfnamefont{C.}~\bibnamefont{{Eichler}}},
  \bibinfo{author}{\bibfnamefont{D.}~\bibnamefont{{Bozyigit}}},
  \bibinfo{author}{\bibfnamefont{C.}~\bibnamefont{{Lang}}},
  \bibinfo{author}{\bibfnamefont{L.}~\bibnamefont{{Steffen}}},
  \bibinfo{author}{\bibfnamefont{J.}~\bibnamefont{{Fink}}}, \bibnamefont{and}
  \bibinfo{author}{\bibfnamefont{A.}~\bibnamefont{{Wallraff}}},
  \bibinfo{journal}{ArXiv e-prints}  (\bibinfo{year}{2010}),
  \eprint{1011.6668}.

\bibitem[{\citenamefont{{Mallet} et~al.}(2010)\citenamefont{{Mallet},
  {Castellanos-Beltran}, {Ku}, {Irwin}, {Vale}, and {Lehnert}}}]{mallet10}
\bibinfo{author}{\bibfnamefont{F.}~\bibnamefont{{Mallet}}},
  \bibinfo{author}{\bibfnamefont{M.}~\bibnamefont{{Castellanos-Beltran}}},
  \bibinfo{author}{\bibfnamefont{H.}~\bibnamefont{{Ku}}},
  \bibinfo{author}{\bibfnamefont{K.}~\bibnamefont{{Irwin}}},
  \bibinfo{author}{\bibfnamefont{L.}~\bibnamefont{{Vale}}}, \bibnamefont{and}
  \bibinfo{author}{\bibfnamefont{K.}~\bibnamefont{{Lehnert}}},
  \bibinfo{journal}{APS Meeting Abstracts} p. \bibinfo{pages}{26014}
  (\bibinfo{year}{2010}).

\bibitem[{\citenamefont{Mariantoni et~al.}()\citenamefont{Mariantoni, Storcz,
  Wilhelm, Oliver, Emmert, Marx, Gross, Christ, and Solano}}]{mariantoni05}
\bibinfo{author}{\bibfnamefont{M.}~\bibnamefont{Mariantoni}},
  \bibinfo{author}{\bibfnamefont{M.~J.} \bibnamefont{Storcz}},
  \bibinfo{author}{\bibfnamefont{F.~K.} \bibnamefont{Wilhelm}},
  \bibinfo{author}{\bibfnamefont{W.~D.} \bibnamefont{Oliver}},
  \bibinfo{author}{\bibfnamefont{A.}~\bibnamefont{Emmert}},
  \bibinfo{author}{\bibfnamefont{A.}~\bibnamefont{Marx}},
  \bibinfo{author}{\bibfnamefont{R.}~\bibnamefont{Gross}},
  \bibinfo{author}{\bibfnamefont{H.}~\bibnamefont{Christ}}, \bibnamefont{and}
  \bibinfo{author}{\bibfnamefont{E.}~\bibnamefont{Solano}},
  \bibinfo{howpublished}{arXiv:cond-mat/0509737}.

\bibitem[{\citenamefont{Mariantoni et~al.}(2010)\citenamefont{Mariantoni,
  Menzel, Deppe, Araque~Caballero, Baust, Niemczyk, Hoffmann, Solano, Marx, and
  Gross}}]{mariantoni10}
\bibinfo{author}{\bibfnamefont{M.}~\bibnamefont{Mariantoni}},
  \bibinfo{author}{\bibfnamefont{E.~P.} \bibnamefont{Menzel}},
  \bibinfo{author}{\bibfnamefont{F.}~\bibnamefont{Deppe}},
  \bibinfo{author}{\bibfnamefont{M.~A.} \bibnamefont{Araque~Caballero}},
  \bibinfo{author}{\bibfnamefont{A.}~\bibnamefont{Baust}},
  \bibinfo{author}{\bibfnamefont{T.}~\bibnamefont{Niemczyk}},
  \bibinfo{author}{\bibfnamefont{E.}~\bibnamefont{Hoffmann}},
  \bibinfo{author}{\bibfnamefont{E.}~\bibnamefont{Solano}},
  \bibinfo{author}{\bibfnamefont{A.}~\bibnamefont{Marx}}, \bibnamefont{and}
  \bibinfo{author}{\bibfnamefont{R.}~\bibnamefont{Gross}},
  \bibinfo{journal}{Phys. Rev. Lett.} \textbf{\bibinfo{volume}{105}},
  \bibinfo{pages}{133601} (\bibinfo{year}{2010}).

\bibitem[{\citenamefont{Menzel et~al.}(2010)\citenamefont{Menzel, Deppe,
  Mariantoni, Araque~Caballero, Baust, Niemczyk, Hoffmann, Marx, Solano, and
  Gross}}]{menzel10}
\bibinfo{author}{\bibfnamefont{E.~P.} \bibnamefont{Menzel}},
  \bibinfo{author}{\bibfnamefont{F.}~\bibnamefont{Deppe}},
  \bibinfo{author}{\bibfnamefont{M.}~\bibnamefont{Mariantoni}},
  \bibinfo{author}{\bibfnamefont{M.~A.} \bibnamefont{Araque~Caballero}},
  \bibinfo{author}{\bibfnamefont{A.}~\bibnamefont{Baust}},
  \bibinfo{author}{\bibfnamefont{T.}~\bibnamefont{Niemczyk}},
  \bibinfo{author}{\bibfnamefont{E.}~\bibnamefont{Hoffmann}},
  \bibinfo{author}{\bibfnamefont{A.}~\bibnamefont{Marx}},
  \bibinfo{author}{\bibfnamefont{E.}~\bibnamefont{Solano}}, \bibnamefont{and}
  \bibinfo{author}{\bibfnamefont{R.}~\bibnamefont{Gross}},
  \bibinfo{journal}{Phys. Rev. Lett.} \textbf{\bibinfo{volume}{105}},
  \bibinfo{pages}{100401} (\bibinfo{year}{2010}).

\bibitem[{\citenamefont{{Knill} et~al.}(2001)\citenamefont{{Knill}, {Laflamme},
  and {Milburn}}}]{knill01}
\bibinfo{author}{\bibfnamefont{E.}~\bibnamefont{{Knill}}},
  \bibinfo{author}{\bibfnamefont{R.}~\bibnamefont{{Laflamme}}},
  \bibnamefont{and} \bibinfo{author}{\bibfnamefont{G.~J.}
  \bibnamefont{{Milburn}}}, \bibinfo{journal}{\nat}
  \textbf{\bibinfo{volume}{409}}, \bibinfo{pages}{46} (\bibinfo{year}{2001}).

\bibitem[{\citenamefont{Romero et~al.}(2009{\natexlab{a}})\citenamefont{Romero,
  Garc\'\i{}a-Ripoll, and Solano}}]{romero09a}
\bibinfo{author}{\bibfnamefont{G.}~\bibnamefont{Romero}},
  \bibinfo{author}{\bibfnamefont{J.~J.} \bibnamefont{Garc\'\i{}a-Ripoll}},
  \bibnamefont{and} \bibinfo{author}{\bibfnamefont{E.}~\bibnamefont{Solano}},
  \bibinfo{journal}{Phys. Rev. Lett.} \textbf{\bibinfo{volume}{102}},
  \bibinfo{pages}{173602} (\bibinfo{year}{2009}{\natexlab{a}}).

\bibitem[{\citenamefont{Romero et~al.}(2009{\natexlab{b}})\citenamefont{Romero,
  Garc{\'\i}≠a-Ripoll, and Solano}}]{romero09b}
\bibinfo{author}{\bibfnamefont{G.}~\bibnamefont{Romero}},
  \bibinfo{author}{\bibfnamefont{J.~J.} \bibnamefont{Garc{\'\i}≠a-Ripoll}},
  \bibnamefont{and} \bibinfo{author}{\bibfnamefont{E.}~\bibnamefont{Solano}},
  \bibinfo{journal}{Physica Scripta} \textbf{\bibinfo{volume}{2009}},
  \bibinfo{pages}{014004} (\bibinfo{year}{2009}{\natexlab{b}}).

\bibitem[{\citenamefont{Helmer et~al.}(2009)\citenamefont{Helmer, Mariantoni,
  Solano, and Marquardt}}]{helmer09}
\bibinfo{author}{\bibfnamefont{F.}~\bibnamefont{Helmer}},
  \bibinfo{author}{\bibfnamefont{M.}~\bibnamefont{Mariantoni}},
  \bibinfo{author}{\bibfnamefont{E.}~\bibnamefont{Solano}}, \bibnamefont{and}
  \bibinfo{author}{\bibfnamefont{F.}~\bibnamefont{Marquardt}},
  \bibinfo{journal}{Phys. Rev. A} \textbf{\bibinfo{volume}{79}},
  \bibinfo{pages}{052115} (\bibinfo{year}{2009}).

\bibitem[{\citenamefont{Pinotsi and Imamoglu}(2008)}]{pinotsi08}
\bibinfo{author}{\bibfnamefont{D.}~\bibnamefont{Pinotsi}} \bibnamefont{and}
  \bibinfo{author}{\bibfnamefont{A.}~\bibnamefont{Imamoglu}},
  \bibinfo{journal}{Phys. Rev. Lett.} \textbf{\bibinfo{volume}{100}},
  \bibinfo{pages}{093603} (\bibinfo{year}{2008}).

\bibitem[{\citenamefont{{Chen} et~al.}(2010)\citenamefont{{Chen}, {Hover},
  {Sendelbach}, {Maurer}, {Merkel}, {Pritchett}, {Wilhelm}, and
  {McDermott}}}]{chen10}
\bibinfo{author}{\bibfnamefont{Y.}~\bibnamefont{{Chen}}},
  \bibinfo{author}{\bibfnamefont{D.}~\bibnamefont{{Hover}}},
  \bibinfo{author}{\bibfnamefont{S.}~\bibnamefont{{Sendelbach}}},
  \bibinfo{author}{\bibfnamefont{L.}~\bibnamefont{{Maurer}}},
  \bibinfo{author}{\bibfnamefont{S.~T.} \bibnamefont{{Merkel}}},
  \bibinfo{author}{\bibfnamefont{E.~J.} \bibnamefont{{Pritchett}}},
  \bibinfo{author}{\bibfnamefont{F.~K.} \bibnamefont{{Wilhelm}}},
  \bibnamefont{and}
  \bibinfo{author}{\bibfnamefont{R.}~\bibnamefont{{McDermott}}},
  \bibinfo{journal}{ArXiv e-prints}  (\bibinfo{year}{2010}),
  \eprint{1011.4329}.

\bibitem[{\citenamefont{Dong et~al.}(2009)\citenamefont{Dong, Gong, Ian, Zhou,
  and Sun}}]{dong09}
\bibinfo{author}{\bibfnamefont{H.}~\bibnamefont{Dong}},
  \bibinfo{author}{\bibfnamefont{Z.~R.} \bibnamefont{Gong}},
  \bibinfo{author}{\bibfnamefont{H.}~\bibnamefont{Ian}},
  \bibinfo{author}{\bibfnamefont{L.}~\bibnamefont{Zhou}}, \bibnamefont{and}
  \bibinfo{author}{\bibfnamefont{C.~P.} \bibnamefont{Sun}},
  \bibinfo{journal}{Phys. Rev. A} \textbf{\bibinfo{volume}{79}},
  \bibinfo{pages}{063847} (\bibinfo{year}{2009}).

\bibitem[{\citenamefont{Shen and Fan}(2005)}]{shen05}
\bibinfo{author}{\bibfnamefont{J.-T.} \bibnamefont{Shen}} \bibnamefont{and}
  \bibinfo{author}{\bibfnamefont{S.}~\bibnamefont{Fan}},
  \bibinfo{journal}{Physical Review Letters} \textbf{\bibinfo{volume}{95}},
  \bibinfo{eid}{213001} (\bibinfo{year}{2005}).

\bibitem[{\citenamefont{Witthaut and S{\o}rensen}(2010)}]{witthaut10}
\bibinfo{author}{\bibfnamefont{D.}~\bibnamefont{Witthaut}} \bibnamefont{and}
  \bibinfo{author}{\bibfnamefont{A.~S.} \bibnamefont{S{\o}rensen}},
  \bibinfo{journal}{New Journal of Physics} \textbf{\bibinfo{volume}{12}},
  \bibinfo{pages}{043052} (\bibinfo{year}{2010}).

\end{thebibliography}

\end{document}